\documentclass[conference]{IEEEtran}
\IEEEoverridecommandlockouts
\usepackage{cite}
\usepackage{amsmath,amssymb,amsfonts}
\usepackage{algorithmic}
\usepackage{graphicx}
\usepackage{textcomp}
\usepackage{xcolor}
\usepackage{booktabs}
\def\BibTeX{{\rm B\kern-.05em{\sc i\kern-.025em b}\kern-.08em
    T\kern-.1667em\lower.7ex\hbox{E}\kern-.125emX}}
\begin{document}

\title{Joint Estimation of DOA and Frequency with Sub-Nyquist Sampling in a Binary Array Radar System\\
\thanks{Identify applicable funding agency here. If none, delete this.}
}
	\author{\IEEEauthorblockN{Zhan Zhang\IEEEauthorrefmark{1}, 
				Ping Wei\IEEEauthorrefmark{1}, Lijuan Deng\IEEEauthorrefmark{1},
				Huaguo Zhang\IEEEauthorrefmark{1}\IEEEauthorrefmark{2}}
		\IEEEauthorblockA{\IEEEauthorrefmark{1}School 
		of Information and Communication Engineering,\\
		University of Electronic Science and Technology of China,
		Chengdu 611731, China\\
        \IEEEauthorrefmark{1}Email: zzhenry15@163.com}
		\IEEEauthorblockA{\IEEEauthorrefmark{2}Science and 
			Technology on Communication Information Security Control Laboratory,\\ 
			Jiaxing, China}}

\maketitle

\begin{abstract}
Recently, several array radar structures 
combined with sub-Nyquist techniques and corresponding algorithms 
have been extensively studied. 
Carrier frequency and direction-of-arrival (DOA) estimations 
of multiple narrow-band signals received by array radars 
at the sub-Nyquist rates 
are considered in this paper. 
We propose  a new sub-Nyquist array radar architecture
(a binary array radar separately connected to a multi-coset structure with $M$ branches)
and an efficient joint estimation algorithm 
which can match frequencies up with corresponding DOAs. 
We further come up with a delay pattern augmenting method,
by which the capability of the number of identifiable signals 
can increase from $M-1$ to $Q-1$ 
($Q$ is extended degrees of freedom).
We further conclude that 
the minimum total sampling rate $2MB$ 
is sufficient to identify $ {K \leq Q-1}$ narrow-band signals of maximum bandwidth $B$ inside.
The effectiveness and performance 
of the estimation algorithm together with the augmenting method
have been verified by simulations.
\end{abstract}


\section{Introduction}
In array radar signal processing, 
joint frequency and direction-of-arrival (DOA) estimation problems 
have received extensive attention. 
Modern signals have relatively high carrier frequencies and are distributed over a comparatively wide spectrum range. 
According to Nyquist Sampling\cite{50afterShannon}, 
samplers will face high conversion speed problem and considerable data pressure,
which facilitates the rapid development of sub-Nyquist sampling techniques \cite{Eldar2014STheory}. 
Therefore, how array radars can jointly estimate frequencies and orientations of signals effectively
at sub-Nyquist Sampling has been extensively studied.

In the literature, many methods such as\cite{Lemma2003JAFEuESPRIT,Wang2012EspritJDF} 
have efficiently achieved 
joint estimations of DOAs and corresponding carrier frequencies at Nyquist sampling. 
However, these methods cannot work well at sub-Nyquist sampling. 
The approaches such as \cite{Zoltowski1995JF2DESSTS,Ariannanda2014CJAFPS} 
overcome this defect and can make joint estimation at sub-Nyquist sampling.
Nevertheless, Zoltowski et al present \cite{Zoltowski1995JF2DESSTS} to decompose the entire spectrum into overlapping sub-band
via such a structure (each antenna followed by a filter bank and a mixer)
and each band has to be examined separately by two paths (a direct path and a delayed path),
which causes more hardware resources and quite slow processing speed.
On the other hand, Ariananda et al \cite{Ariannanda2014CJAFPS} put forward
to connect a multi-coset structure \cite{Eldar2009MCoset} to each antenna
of the minimum redundancy linear array (MRA) \cite{Moffet2003MRA} 
to implement joint estimation with sub-Nyquist sampling.
This array arrangement also causes high consumption.
Recently, Liu et al come up with \cite{Liu2017JDFESMSS,Liu2018JDFESSAS} 
to jointly estimate frequencies and DOAs with sub-Nyquist sampling for more sources than sensors.
However, these methods also bring high hardware pressure;
the total sampling rate, 
i.e., the sum of sampling rates for all channels, 
is fairly high 
even though the sampling rate of each channel is lower than the Nyquist sampling rate.

To overcome the problems of high hardware resource consumption and high total sampling rate, 
Kumar et al have suggested a simple sub-Nyquist sampling architecture\cite{Kumar2014ESRSBR,Kumar2015CFDE} 
together with the corresponding algorithm. 
This architecture is that each receiver has only two paths, 
a direct path and a delayed path, followed by an ADC separately. 
The corresponding algorithm is based on 
MUSIC algorithm\cite{Schmidt1986MUSIC} and ESPRIT algorithm\cite{Roy1986Esprits}, 
which can realize joint estimations of DOAs and frequencies at a fairly low total sampling rate. 
Due to the limitation of Kruskal rank\cite{Kruskal1977}, 
in order to avoid the blurring problem of frequency and DOA estimation, 
this algorithm cannot be applied the in uniform linear array (ULA), 
but in circular shape array\cite{Kumar2014ESRSBR} and rectangular nested array\cite{Kumar2015CFDE}, 
which illustrates some certain limitations for the array layout. 
Moreover, based on the generalized eigendecomposition of the matrix beam, 
its frequency estimation is too sensitive to noise. 
Thus, a two-dimensional multiresolution algorithm has to be applied to solve performance problems.

The paper aims to
to achieve joint estimations of frequencies and corresponding DOAs effectively 
at a very low total sampling rate 
in an array radar system, 
where the number of array elements is fewer than that of signals. 
We propose a binary array radar structure and corresponding joint estimation algorithm with high efficiency, 
which can automatically pair frequencies and DOAs 
to overcome the ambiguity problem\cite{Kumar2014ESRSBR}.
Moreover, an augmenting method similar to literature \cite{Liu2018JDFESSAS} is presented, 
which can be applied to estimate more signal parameters with fewer channels. 

The paper is organized as follows: 
in the next section, we introduce the scenario 
together with signal model and present a binary array radar architecture. 
In section III, we describe our algorithm as well as an expansion of time-delay manifold method, 
and give the boundary of the minimal total sampling rate. 
Section IV shows simulation results and offers some further discussion. 
Finally, section V comes to the conclusion.

\section{Signal Model and Proposed Array Structure}
In this section, we first describe the scenario 
including the signal model and some specific assumptions. 
Then a binary array radar structure is introduced, 
and corresponding array received signal model is given.

\subsection{Signal Model}
Consider such a scenario where $K$ uncorrelated, narrow-band, far-field signals 
spreading over a very wide spectrum range from different orientations
are received by a radar array. 
These $K$ signals, considered as time domain wide multiband signals (MBS), 
i.e., many disjoint narrow-band signals within a wide-band spectrum, 
are denoted as $x\left( t \right)$ and can be expressed as
\begin{equation}\label{eq-x(t)}
x\left( t \right) = \sum\limits_{k = 1}^K {{s_k}\left( t \right){e^{j2\pi {f_k}t}}} \
\end{equation}
Then, the time delay expression of $x\left( t \right)$ with time difference $\tau$ can be written as
\begin{equation}\label{eq-x(t-tao)}
\begin{aligned}
x\left( {t - \tau } \right) &= \sum\limits_{k = 1}^K {{s_k}\left( {t - \tau } \right){e^{j2\pi {f_k}\left( {t - \tau } \right)}}}\\  
&\approx \sum\limits_{k = 1}^K {{s_k}\left( t \right){e^{j2\pi {f_k}\left( {t - \tau } \right)}}} 
= \sum\limits_{k = 1}^K {{{\tilde s}_k}\left( t \right){e^{ - j2\pi {f_k}\tau }}} 
\end{aligned}
\end{equation}
where ${s_k}\left( t \right),{\tilde s_k}\left( t \right), k = \left\{ {1,2, \dots ,K} \right\}$ 
denote the ${k^{th}}$ baseband signal and corresponding modulated signal with carrier frequency ${f_k}$, respectively.
The reason of the approximation in \eqref{eq-x(t-tao)} is the narrow-band assumption. 
We further make other assumptions similar to\cite{Kumar2014ESRSBR} on the above signal model:
\begin{itemize}\label{assumption}
	\item The baseband signals $\left\{ {{s_k}\left( t \right)} \right\}_{k = 1}^K$ 
	are supposed to be mutually orthogonal and are bandlimited to
	${B_k} = \left[ { - {{{B_k}} \mathord{\left/{\vphantom {{{B_k}} {2,{{{B_k}} \mathord{\left/{\vphantom {{{B_k}} 2}} \right.\kern-\nulldelimiterspace} 2}}}} \right.\kern-\nulldelimiterspace} {2,{{{B_k}} \mathord{\left/{\vphantom {{{B_k}} 2}} \right.\kern-\nulldelimiterspace} 2}}}} \right],k = \left\{ {1,2,\dots,K} \right\}$. 
	We further assume that 
$\forall k = \left\{ {1,2, \dots ,K} \right\},{B_k} \le B$ . 
	\item The MBS $x\left( t \right)$ is assumed to be bandlimited to 
	${\mathcal F} = \left[ {0,{1 	\mathord{\left/{\vphantom {1 T}} \right.
				\kern-\nulldelimiterspace} T}} \right]$
	and $T$ denotes the Nyquist sampling interval of  $x\left( t \right)$. 
	We further assume that 
	${1 \mathord{\left/{\vphantom {1 T}} \right.\kern-\nulldelimiterspace} T} \gg B$, 
	that is to say, the total bandwidth of signals is far less than the total spectrum range.
	\item Assume that the information bands do not overlap, 
	i.e., $\{ {I( {{{\tilde S}_i}\left( f \right)} ) \cap I( {{{\tilde S}_j}\left( f \right)} ) = \mathbf{\emptyset} :i,j \in \left\{ {1,2, \dots ,K} \right\}} \}$
	where ${{\tilde S}_i}\left( f \right)$ denotes the Fourier transform of modulated signal 
	${s_i}\left( t \right){e^{j2\pi {f_i}t}}$ and $I( {{\tilde S}_i}\left( f \right))$ 
	represents the support region of ${{\tilde S}_i}\left( f \right)$. 
	This assumption satisfies that the carrier frequencies $\left\{ {{f_i}} \right\}_{i = 1}^K$ 
	are distinct and corresponding signal spectral bands do not overlap.
\end{itemize}

\subsection{Proposed Array Architecture}
\begin{figure}[htbp]
	\centerline{\includegraphics[width=8.0cm]{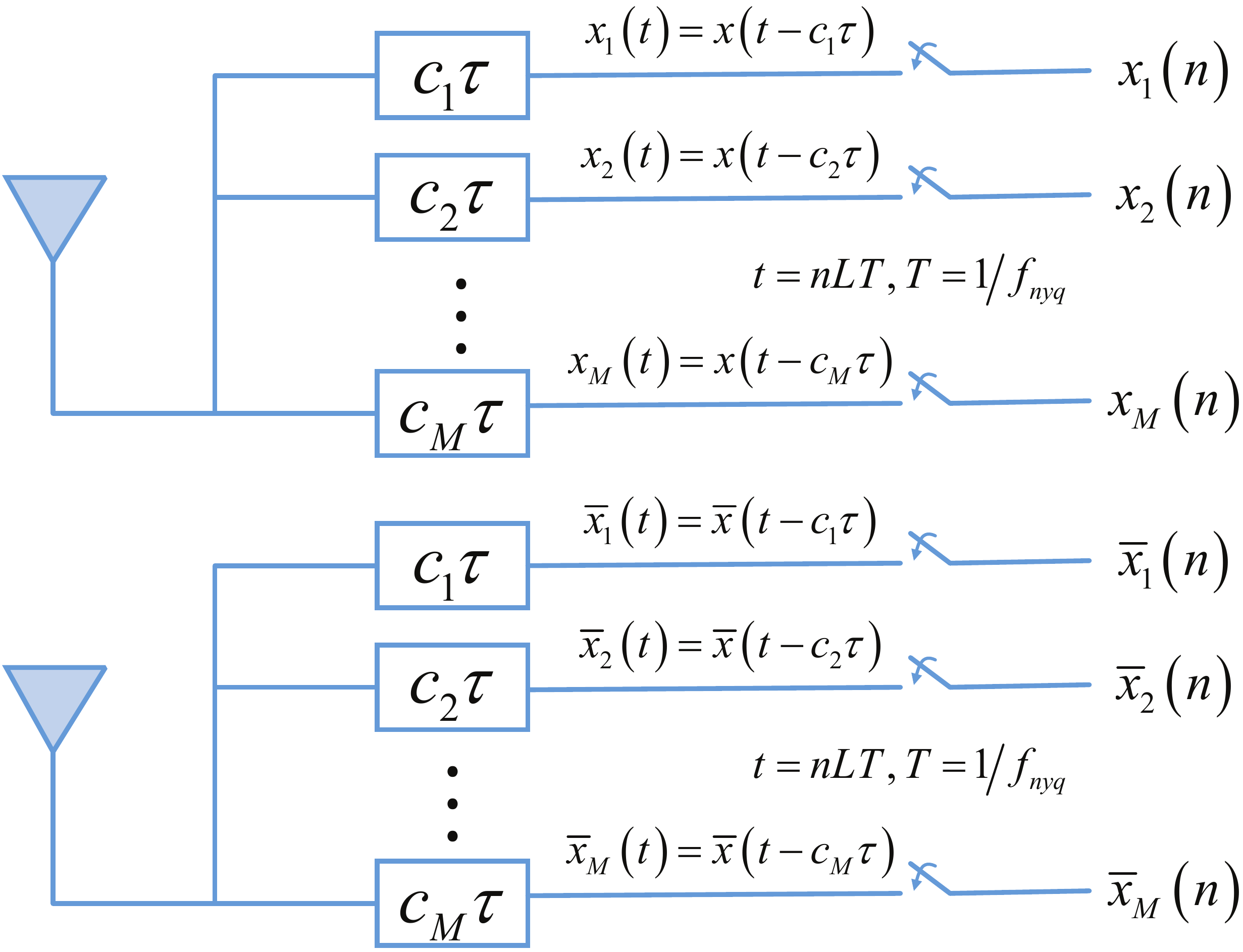}}
	\caption{Proposed Array Radar Structure (Binary Array)}
	\label{fig1}
\end{figure}
Fig.1 depicts the receiving array radar structure proposed in this paper. 
This array structure has only two array elements, 
and the first one is considered as the reference element. 
The distance between the two elements is denoted as $d$, 
and $d$ is equal to the half-wavelength of the signal whose frequency is 
the Nyquist sampling rate $f_{nyq}$. 
Each array element connects separately to an identical multi-coset structure 
including $M$ branches followed by samplers at $LT$ sampling intervals, 
where $L$ and $T$ denote the sampling rate reduction factor 
and the Nyquist sampling interval, respectively. 
Denote the minimal delay unit as $\tau$ 
and the delay coefficients of these $M$  channels are denoted as 
$C = \left[ {{c_1}, \ldots ,{c_M}} \right],0 = {c_1} <  \cdots  < {c_M}$. 
The delay pattern can be similarly set up to the array pattern of a sparse array, 
such as MRAs \cite{Moffet2003MRA}, coprime arrays.

We can express the output signal of the ${m^{th}}$ path of the reference element as ${x_m}\left( t \right)$ 
and further define the signal model of the other array element as the form $\overline {\left( \cdot \right)}$. 
Therefore, the outputs of the ${m^{th}}$ path of the reference element and the other one  ${x_m}\left( t \right)$, ${\bar x_m}\left( t \right)$, can be expressed as
\begin{equation}\label{eq-x_m}
\renewcommand{\arraystretch}{2}
\begin{array}{l}
{x_m}\left( t \right) = \sum\limits_{k = 1}^K {{{\tilde s}_k}\left( t \right){e^{ - j{\omega _k}{c_m}}}}  + {n_m}\left( t \right)\\
{{\bar x}_m}\left( t \right) = \sum\limits_{k = 1}^K {{{\tilde s}_k}\left( t \right){e^{ - j{\omega _k}{c_m}}}{e^{ - j{\phi _k}}}}  + {{\bar n}_m}\left( t \right)
\end{array}
\end{equation}
where ${\omega _k} \buildrel \Delta \over = 2\pi {f_k}\tau$,
${\phi _k} \buildrel \Delta \over = {{{\rm{2}}\pi d{f_k}\sin \left( {{\theta _k}} \right)} \mathord{\left/
		{\vphantom {{{\rm{2}}\pi d{f_k}\sin \left( {{\theta _k}} \right)} c}} \right.
		\kern-\nulldelimiterspace} c}$,
and ${n_m}\left( t \right),{\bar n_m}\left( t \right)$ 
are additive noise of corresponding paths.
${\theta _k}$ is the DOA of the ${k^{th}}$ signal. 
${\omega _k}$ is the unit phase difference caused by the delay unit, 
including only frequency information ${f_k}$;
${\phi _k}$ is the unit phase difference derived from the array structure, 
including joint frequency and angle information $\{f_k, \theta _k\}$.
Moreover, the additive noise of all channels obeys zero-mean Gaussian distribution 
with variance ${\sigma ^2}$ and is statistically independent of signals.

Consider the received signals of all channels written as matrix form
\begin{equation}\label{eq-x_all-matrix}
\renewcommand{\arraystretch}{1.2}
{{\bf{x}}_{all}} = \left[ {\begin{array}{*{20}{c}}
	{{\bf{x}}\left( t \right)}\\
	{{\bf{\bar x}}\left( t \right)}
	\end{array}} \right] = \left[ {\begin{array}{*{20}{c}}
	{{{\bf{A}}_t}}\\
	{{{\bf{A}}_t}{{\bf{D}}_\phi }}
	\end{array}} \right]{\bf{s}}\left( t \right) + \left[ {\begin{array}{*{20}{c}}
	{{\bf{n}}\left( t \right)}\\
	{{\bf{\bar n}}\left( t \right)}
	\end{array}} \right]
\end{equation}
where  ${{\bf{A}}_t}\left( {\bf{\omega }} \right) = {\left[ {\begin{array}{*{20}{c}}
		{{{\bf{a}}_t}\left( {{\omega _1}} \right)}& \cdots &{{{\bf{a}}_t}\left( {{\omega _K}} \right)}
		\end{array}} \right]_{M \times K}}$ 
is time-delay manifold matrix including only the frequency information, 
${{\bf{D}}_\phi}= diag\left( {{e^{ - j{\phi _{1}}}},\dots,{e^{ - j{\phi _{K}}}}} \right)$
is a diagonal matrix including frequency and DOA information, 
and ${\bf{n}}\left( t \right),{\bf{\bar n}}\left( t \right)$ are the matrices of the noise, 
whose rows represent an additive noise of the different paths.

\section{Proposed Approach} 
In this section, we put forward an approach 
to identify as many signals as possible at sub-Nyquist sampling in a binary array radar struture.
This approach contains a joint estimation algorithm 
and an expansion method for increasing the number of identifiable signals.

\subsection{Frequency Domain and Time Domain Analyses}\label{FDAnalysis}
According to\cite{Kumar2014ESRSBR}, at sub-Nyquist structures, 
time domain data is typically converted to frequency domain data for analysis. 
Therefore, the discrete time Fourier transform of sampled signal of the $m^{th}$ path
at ${f_s} = {{{f_{nyq}}} \mathord{\left/{\vphantom {{{f_{nyq}}} L}} \right.
		\kern-\nulldelimiterspace} L}$
sampling rate can be expressed as
\begin{equation}\label{eq-X1_m}
\renewcommand{\arraystretch}{2}
\begin{array}{l}
{X_m}\left( {{e^{j2\pi fT}}} \right) = \sum\limits_{k = 1}^K {{e^{ - j{\omega _k}{c_m}}}S_k^p\left( f \right)}  + N_m^p\left( f \right)\\
{{\bar X}_m}\left( {{e^{j2\pi fT}}} \right) = \sum\limits_{k = 1}^K {{e^{ - j{\omega _k}{c_m}}}{e^{ - j{\phi _k}}}S_k^p\left( f \right)}  + \bar N_m^p\left( f \right)
\end{array}
\end{equation}
where $S_k^p\left( f \right)=\sum_{i \in \mathbb{Z}} {{{\tilde S}_k}\left( {f - i{f_s}} \right)},k \in \left\{ {1,2,...K} \right\},f \in \left[ {0,{f_{sub}}} \right)$ 
denotes the aliased spectrum of ${k^{th}}$ signal, 
and $N_m^p\left( {{f_s}} \right) = \sum_{i \in \mathbb{Z} } {{N_m}\left( {f - i{f_s}} \right)} ,m \in \left\{ {1,2,...,M} \right\}$ represent the aliased noise spectrum. 
Thus, The spectral data of all channels can be reformulated as matrix form
\begin{equation}\label{eq-X_all-matrix}
\renewcommand{\arraystretch}{1.2}
{{\bf{X}}_{all}} = \left[ {\begin{array}{*{20}{c}}
	{{\bf{X}}\left( f \right)}\\
	{{\bf{\bar X}}\left( f \right)}
	\end{array}} \right] = \left[ {\begin{array}{*{20}{c}}
	{{{\bf{A}}_t}}\\
	{{{\bf{A}}_t}{{\bf{D}}_\phi }}
	\end{array}} \right]{\bf{S}}\left( f \right) + \left[ {\begin{array}{*{20}{c}}
	{{\bf{N}}\left( f \right)}\\
	{{\bf{\bar N}}\left( f \right)}
	\end{array}} \right]
\end{equation}
where ${\bf{S}}\left( f \right) = {\left[ {S_1^p\left( f \right), S_2^p\left( f \right), \ldots ,S_K^p\left( f \right)} \right]^T}$
and
${\bf{N}}\left( f \right) = {\left[ {N_1^p\left( f \right),N_2^p\left( f \right), \ldots ,N_M^p\left( f \right)} \right]^T}$.
We form the following covariance matrix
\begin{equation}\label{eq-R_X_all}
\renewcommand{\arraystretch}{1.2}
\begin{aligned}
{{\bf{R}}_{{X_{all}}}} &= E\left\{ {{{\bf{X}}_{all}}{\bf{X}}_{all}^H} \right\} = \left[ {\begin{array}{*{20}{c}}
	{{{\bf{R}}_{XX}}}&{{{\bf{R}}_{X\bar X}}}\\
	{{{\bf{R}}_{\bar XX}}}&{{{\bf{R}}_{\bar X\bar X}}}
	\end{array}} \right]\\
&= \left[ {\begin{array}{*{20}{c}}
	{{{\bf{A}}_t}}\\
	{{{\bf{A}}_t}{{\bf{D}}_\phi }}
	\end{array}} \right]{\bf{W}}{\left[ {\begin{array}{*{20}{c}}
		{{{\bf{A}}_t}}\\
		{{{\bf{A}}_t}{{\bf{D}}_\phi }}
		\end{array}} \right]^H} + {L\sigma ^2}{{\bf{I}}_{2M}}\\
&= {\bf{AW}}{{\bf{A}}^H} + {L\sigma ^2}{{\bf{I}}_{2M}}
\end{aligned}
\end{equation}
where ${\bf{A}} = \left[ {{\bf{a}}\left( {{f_1},{\phi _1}} \right),{\bf{a}}\left( {{f_2},{\phi _2}} \right), \ldots ,{\bf{a}}\left( {{f_k},{\phi _K}} \right)} \right]$
and ${\bf{a}}\left( {{f_k},{\phi _k}} \right)$ 
contains the information of frequency and corresponding DOA of the ${k^{th}}$ signal; 
${\bf{W}} = E\left\{ {{\bf{S}}\left( f \right){{\bf{S}}^H}\left( f \right)} \right\} 
= diag\left({{W_1}},\dots,{{W_K}}\right)$
and ${W_k}$ denotes the ${k^{th}}$ signal power. 
Notice that noise power goes up to $L\sigma^2$, 
because the noise power spectrum is folded under sub-Nyquist sampling. 

\subsection{Proposed Algorithm for Estimation of Carrier Frequency and DOAs}\label{JDF4BA}
Due to the diagnose matrix ${{\bf{D}}_\phi }$, 
we can exchange the positions of ${{\bf{D}}_\phi }$ and ${\bf{W}}$ 
in the matrix block in \eqref{eq-R_X_all},
and the correlation matrix ${{\bf{R}}_{{X_{all}}}}$ becomes the following form
\begin{equation}\label{eq-R_X_all_C}
\renewcommand{\arraystretch}{1.5}
{{\bf{R}}_{{X_{all}}}} = \left[ {\begin{array}{*{20}{c}}
	{{{\bf{A}}_t}{\bf{WA}}_t^H}&{{{\bf{A}}_t}{\bf{WD}}_\phi ^H{\bf{A}}_t^H}\\
 {{{\bf{A}}_t}{\bf{WD}}_\phi {\bf{A}}_t^H}&{{{\bf{A}}_t}{\bf{WA}}_t^H}
	\end{array}} \right] + {L\sigma ^2}{{\bf{I}}_{2M}}
\end{equation}

Notice that the two matrix blocks ${{\bf{R}}_{XX}}$,${{\bf{R}}_{\bar X\bar X}}$ on the diagonal of the correlation matrix ${{\bf{R}}_{{X_{all}}}}$ in \eqref{eq-R_X_all_C} have the same form i.e.,
${{\bf{A}}_t}{\bf{WA}}_t^H{\rm{ + }}{L\sigma ^2}{{\bf{I}}_M}$, which contains only frequency information. 
According to the MUSIC algorithm\cite{Schmidt1986MUSIC}, 
Use eigen-decomposition on the matrix blocks to obtain corresponding matrices ${\bf{\hat G}}$  
composed of eigenvectors corresponding to the noise subspace. 
The scanning function of peseudo spectrum can be written as follow
\begin{equation}\label{eq-Pmusic(f)}
{\hat P}\left( \omega  \right) = \frac{1}{{{\bf{a}}_t^H\left( \omega  \right){\bf{\hat G}}{{{\bf{\hat G}}}^H}{{\bf{a}}_t}\left( \omega  \right)}}
\end{equation} 
As a result, 
use twice MUSIC algorithm on the two matrix blocks to obtain two Pseudo spectrums, 
and estimate carrier frequencies twice. 
Average two frequency estimatons, 
and the final estimates of the frequencies  $\{ {{{\hat f}_k}}\}_{k = 1}^K$can be obtained.

According to \eqref{eq-R_X_all}, 
${\bf{A}}$ contains all information of frequencies and corresponding DOAs, 
which have been paired and cannot become ambiguous. 
Use eigen-decomposition on the correlation matrix ${{\bf{R}}_{{X_{all}}}}$ to obtain corresponding matrix ${\bf{\hat U}}$  
composed of eigenvectors corresponding to the noise subspace. 
Utilize the following scanning funtion to obtain corresponding DOAs
\begin{equation}\label{eq-Pmusic(phi)}
{\hat P}\left( {{{\hat f }_k},\phi } \right) = \frac{1}{{{{\bf{a}}^H}\left( {{{\hat f }_k},\phi } \right){\bf{\hat U}}{{{\bf{\hat U}}}^H}{\bf{a}}\left( {{{\hat f }_k},\phi } \right)}}
\end{equation}
where $ k = \left\{ {1,2, \dots ,K} \right\}$. 
Based on MUSIC algorithm, 
this scanning function \eqref{eq-Pmusic(phi)} 
is combined with frequency estimation $\{ {{{\hat f}_k}}\}_{k = 1}^K$ 
to obtain the corresponding DOA estimation. 
This complete joint estimation algorithm is named after JDF4BA and its specific processing steps are as follows:
\begin{table}[htbp]\label{tab1}
	\caption{Algorithm JDF4BA}
	\footnotesize
	\centering
	\begin{tabular}{cp{7.0cm}p{0.1cm}}
		\toprule
		{Step} &{Operation} \\
		\midrule
		1)&Calculate  covariance matrix ${{\bf{R}}_{{X_{all}}}}$ according to \eqref{eq-R_X_all};\\
		2)&Eigen Decompose matrix blocks ${{{\bf{R}}_{XX}}}$, ${{{\bf{R}}_{\bar X\bar X}}}$ 
		to get ${\bf{\hat G}}$;\\
		3)&Use \eqref{eq-Pmusic(f)} to estimate the frequencies twice;\\
		4)&Average the two frequency estimation to acquire a final frequency estimation $\{ {{{\hat f}_k}}\}_{k = 1}^K$;\\
		5)&Eigen Decompose on ${{\bf{R}}_{{X_{all}}}}$ to obtain ${\bf{\hat U}}$;\\
		6)&According to $\{ {{{\hat f}_k}}\}_{k = 1}^K$, orederly utilize \eqref{eq-Pmusic(phi)} to get joint estimation $\{ {{{\hat f}_k},{{\hat \theta }_k}} \}_{k = 1}^K$, notice $K<M$.\\
		\bottomrule
	\end{tabular}
\end{table}

\subsection{Expansion of Time-Delay Manifold}\label{ETM}
Since the JDF4BA algorithm utilizes the orthogonal characteristic 
between the noise subspace and the manifold vector, 
JDF4BA algorithm can estimate up to $M - 1$ signals. 
The aim of this subsection is to achieve 
the breakthrough of the number of identifiable signals on such a binary array radar
i.e., the more number of  identifiable signals.

According to \eqref{eq-R_X_all_C}, 
we first make column vectorization on 4 matrix blocks of ${{\bf{R}}_{{X_{all}}}}$,
and the outputs are expressed as
\begin{equation}\label{eq-r_X}
\renewcommand{\arraystretch}{1.3}
\begin{array}{l}
{{{\bf{r}}_{XX}} = vec\left( {{{\bf{R}}_{XX}}} \right) = \left( {{\bf{A}}_t^* \odot {{\bf{A}}_t}} \right){\bf{w}} + {L\sigma ^2}{{\bf{i}}_M}}\\
{{{\bf{r}}_{X\bar X}} = vec\left( {{{\bf{R}}_{X\bar X}}} \right) = \left( {{\bf{A}}_t^* \odot {{\bf{A}}_t}} \right){\bf{D}}_\phi ^*{\bf{w}}}\\
{{{\bf{r}}_{\bar XX}} = vec\left( {{{\bf{R}}_{\bar XX}}} \right) = \left( {{\bf{A}}_t^* \odot {{\bf{A}}_t}} \right){{\bf{D}}_\phi }{\bf{w}}}\\
{{{\bf{r}}_{\bar X\bar X}} = vec\left( {{{\bf{R}}_{\bar X\bar X}}} \right) = \left( {{\bf{A}}_t^* \odot {{\bf{A}}_t}} \right){\bf{w}} + {L\sigma ^2}{{\bf{i}}_M}}
\end{array}
\end{equation}
where ${\bf{w}} = {\left[ {{W_1}, \ldots ,{W_K}} \right]^T}$
and ${{\bf{i}}_{M}} = vec( {{\bf{I}}_{M}})$. 
$\odot$, $vec(\cdot)$ represent Khatri-Rao product and column vectorization, respectively.
The Khatri-Rao product of the manifold matrix 
is actually a virtual extension of the manifold matrix. 
Hence, design matrix transformation operator as ${\bf{\Xi }}$ to rearrange virtual manifold matrix 
and obtain the continuous manifold matrix ${{\bf{A}}_t^c}$, 
shown as follow
\begin{equation}\label{eq-A_c}
{{\bf{A}}_t^c} = {\bf{\Xi }}\left( {{\bf{A}}_t^ *  \odot {{\bf{A}}_t}} \right) = {\left[ {\begin{array}{*{20}{c}}
		{{{\bf{a}}_t^c}\left( {{\omega _1}} \right)}& \cdots &{{{\bf{a}}_t^c}\left( {{\omega _K}} \right)}
		\end{array}} \right]}
\end{equation}
where ${\bf{a}}_t^c\left( {{\omega _k}} \right) = {\left[ {{e^{j{\omega _k}\left( {Q - 1} \right)}}, \ldots ,{e^{ - j{\omega _k}\left( {Q - 1} \right)}}} \right]^T}$. 
$Q$ denotes the degree of freedom of expansion. 
Therefore, via matrix transformation operator, we can obtain
\begin{equation}
\renewcommand{\arraystretch}{1.3}
\begin{array}{l}\label{eq-z_X^c}
{\bf{z}}_{XX}^c = {\bf{A}}_t^c{\bf{w}} + {L\sigma ^2}{\bf{\Xi }}{{\bf{i}}_M}\\
{\bf{z}}_{X\bar X}^c = {\bf{A}}_t^c{\bf{D}}_\phi ^*{\bf{w}}\\
{\bf{z}}_{\bar XX}^c = {\bf{A}}_t^c{{\bf{D}}_\phi }{\bf{w}}\\
{\bf{z}}_{\bar X\bar X}^c = {\bf{A}}_t^c{\bf{w}} + {L\sigma ^2}{\bf{\Xi }}{{\bf{i}}_M}
\end{array}
\end{equation}

According to\cite{Liu2018JDFESSAS}, define the extraction matrix as 
${{\bf{\Gamma }}_i} = {\left[ {{{\bf{0}}_{Q \times \left( {i - 1} \right)}},{{\bf{I}}_Q},{{\bf{0}}_{Q \times \left( {Q - i} \right)}}} \right]_{Q \times \left( {2Q - 1} \right)}},\;i = \left\{ {1,2, \dots ,Q} \right\}$ 
and carry out the following process
\begin{equation}
\renewcommand{\arraystretch}{1.3}
\begin{array}{l}\label{eq-z_Xi^c}
{{\bf{z}}_{XXi}^c = {{\bf{\Gamma }}_i}{\bf{z}}_{XX}^c}\\
{{\bf{z}}_{X\bar Xi}^c = {{\bf{\Gamma }}_i}{\bf{z}}_{X\bar X}^c}\\
{{\bf{z}}_{\bar XXi}^c = {{\bf{\Gamma }}_i}{\bf{z}}_{\bar XX}^c}\\
{{\bf{z}}_{\bar X\bar Xi}^c = {{\bf{\Gamma }}_i}{\bf{z}}_{\bar X\bar X}^c}
\end{array}
\end{equation}

Rearrange the above expressions and obtain new correlation matrix blocks as follows
\begin{equation}
\renewcommand{\arraystretch}{1.5}
\begin{array}{l}\label{eq-R_X^v}
{\bf{R}}_{XX}^v = \left[ {\begin{array}{*{20}{c}}
	{{\bf{z}}_{XXQ}^c}& \cdots &{{\bf{z}}_{XX{\rm{1}}}^c}
	\end{array}} \right] \\
   \phantom{{\bf{R}}_{XX}^v }= {\bf{A}}_t^v{\bf{W}}{({\bf{A}}_t^v)^H} + {L\sigma ^2}{{\bf{I}}_Q}\\
{\bf{R}}_{X\bar X}^v = \left[ {\begin{array}{*{20}{c}}
	{{\bf{z}}_{X\bar XQ}^c}& \cdots &{{\bf{z}}_{X\bar X{\rm{1}}}^c}
	\end{array}} \right] \\
\phantom{{\bf{R}}_{XX}^v }= {\bf{A}}_t^v{\bf{WD}}_\phi ^H{({\bf{A}}_t^v)^H}\\
{\bf{R}}_{\bar XX}^v = \left[ {\begin{array}{*{20}{c}}
	{{\bf{z}}_{\bar XXQ}^c}& \cdots &{{\bf{z}}_{\bar XX{\rm{1}}}^c}
	\end{array}} \right] \\
\phantom{{\bf{R}}_{XX}^v }= {\bf{A}}_t^v{\bf{W}}{{\bf{D}}_\phi }{({\bf{A}}_t^v)^H}\\
{\bf{R}}_{\bar X\bar X}^v = \left[ {\begin{array}{*{20}{c}}
	{{\bf{z}}_{\bar X\bar XQ}^c}& \cdots &{{\bf{z}}_{\bar X\bar X{\rm{1}}}^c}
	\end{array}} \right] \\
\phantom{{\bf{R}}_{XX}^v }= {\bf{A}}_t^v{\bf{W}}{({\bf{A}}_t^v)^H} + {L\sigma ^2}{{\bf{I}}_Q}
\end{array}
\end{equation}
where
${\bf{A}}_t^v = \left[ {{\bf{a}}_t^v\left( {{\omega _1}} \right), \dots ,{\bf{a}}_t^v\left( {{\omega _K}} \right)} \right]$  denotes the virtual manifold matrix,  
and   ${\bf{a}}_t^v\left( {{\omega _k}} \right) = \left[ {{e^{ - j{\omega _k}0}}, \dots ,{e^{ - j{\omega _k}\left( {Q - 1} \right)}}} \right]$ represents the virtual manifold vector. Collect the above correlation matrix blocks and form a new correlation matrix
\begin{equation}\label{eq-R_X_all_v}
\renewcommand{\arraystretch}{1.5}
{\bf{R}}_{{X_{all}}}^v = \left[ {\begin{array}{*{20}{c}}
  {{\bf{A}}_t^v}{\bf{W}}({{\bf{A}}_t^v})^H& {{\bf{A}}_t^v}{\bf{WD}}_\phi ^H({{\bf{A}}_t^v})^H\\
{{\bf{A}}_t^v}{\bf{WD}}_\phi({{\bf{A}}_t^v})^H& {{\bf{A}}_t^v}{\bf{W}}({{\bf{A}}_t^v})^H
	\end{array}} \right] + {L\sigma ^2}{{\bf{I}}_{2Q}}
\end{equation}

It can be seen 
that the structure of the new correlation matrix 
is actually similar to that of  the correlation matrix ${{\bf{R}}_{{X_{all}}}}$ in \eqref{eq-R_X_all_C}, 
and frequencies and DOAs in the new manifold matrix have been already matched up,
and the dimension of the correlation matrix increases from $M$ to $Q$.
Thus, we can use JDF4BA algorithm on these new correlation matrix ${\bf{R}}_{{X_{all}}}^v$ 
to identify $Q-1$ signals,
and this expansion of time-delay manifold method is named after ETM method.
The specific steps for JDF4BA algorithm combined with ETM method to identify signals is shown
in TABLE II below
\begin{table}[htbp]\label{tab2}
	\caption{Algorithm JDF4BA with ETM}\label{JDF4BA via ETM}
	\footnotesize
	\centering
	\begin{tabular}{cp{7.0cm}p{0.1cm}}
		\toprule
		{Step} &{Operation} \\
		\midrule
	1)&Calculate  covariance matrix ${{\bf{R}}_{{X_{all}}}}$ according to \eqref{eq-R_X_all};\\
	2)&Make column vectorization on matrix blocks of ${{\bf{R}}_{{X_{all}}}}$ according to \eqref{eq-r_X};\\
	3)&Utilize \eqref{eq-A_c} to select the continuous manifold matrix and obtain \eqref{eq-z_X^c};\\
	4)&Use extraction ${{\bf{\Gamma }}_i}$ and rearrange vectors according to \eqref{eq-R_X^v};\\
	5)&Collect the new matrix blocks to obtain new correlation matrix  ${\bf{R}}_{{X_{all}}}^v$ according to \eqref{eq-R_X_all_v};\\
    6)&Use algorithm JDF4BA on new correlation matrix ${\bf{R}}_{{X_{all}}}^v$ to get joint estimation $\{ {{{\hat f}_k},{{\hat \theta }_k}} \}_{k = 1}^K$, notice $K<Q$.\\   
		\bottomrule
	\end{tabular}
\end{table}
\subsection{Analyses of sampling rate reduction factor and the number of identifiable signals}\label{Analyses}

With the proposed method on the binary array, the joint estimation technique is effective if 
\begin{enumerate}
	\item[i.] $Q \ge K + 1$
	\item[ii.] $L \le 1/BT$.
\end{enumerate}
The algorithm JDF4BA based on MUSIC algorithm 
can estimate the carrier frequencies and corresponding DOAs of MBS signal, 
by utilizing the noise subspace. 
Therefore, the dimension $Q$ of expansion of time-delay manifold 
has to be larger than the number $K$ of signals.
For the second condition, 
due to assumptions from \ref{assumption}, 
${f_s}$ should  satisfy that the periodic spectrums i.e.,
$S_k^p\left( f \right)=\sum_{i \in \mathbb{Z}} {{{\tilde S}_k}\left( {f - i{f_s}} \right)},k \in \left\{ {1,2,...K} \right\},f \in \left[ {0,{f_{nyq}}} \right)$ 
do not alias. 
This is possible if ${f_s} \ge B$ or $L \le 1/BT$, 
the maximum bandwidth is assumed to be $B$ according to the assumption.

\section{Simulation}
\begin{figure*}[htbp]
	\centerline{\includegraphics{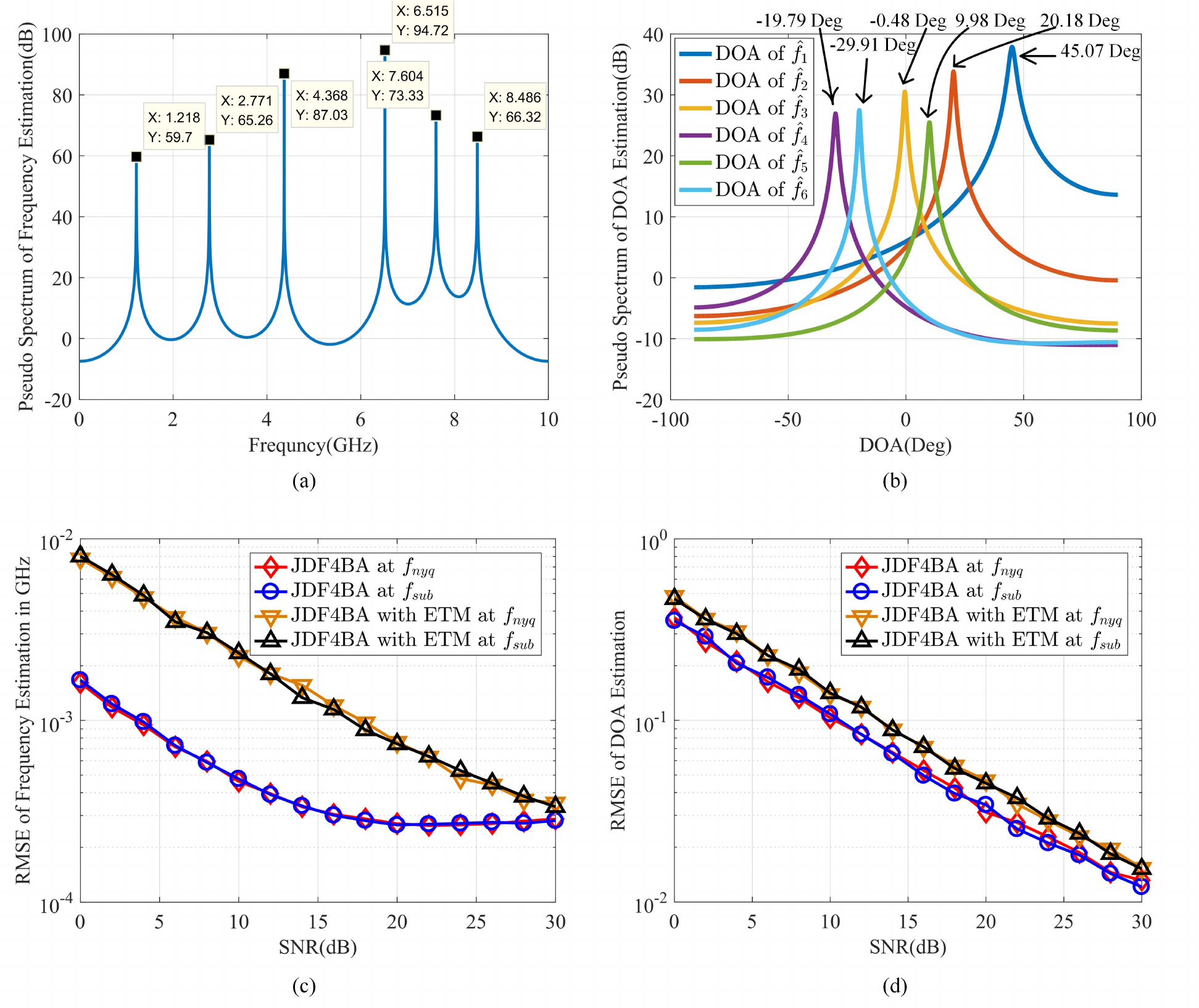}}
	\caption{ Pseudo spectrum via algorithm JDF4BA with ETM (the bandwidth of each signal $B = 25 \text{MHz}$, SNR $=$ 10dB, and the sub-Nyquist sampling rate  $f_{sub} = 25 \text{MHz}$, i.e., the sampling rate reduction factor $L=400$): (a) Pseudo spectrum of frequency estimation according to \eqref{eq-Pmusic(f)}; (b) Pseudo spectrum of DOA estimation \eqref{eq-Pmusic(phi)}. Performance comparison (all are  complex-valuded sinusoid signals, the Nyquist sampling rate $f_{nyq} = 10 \text{GHz}$, and the sub-Nyquist sampling rate $f_{sub} = 100 \text{MHz}$, i.e., the sampling rate reduction factor $L=100$): (c) Frequency estimation performance comparison; (d) DOA estimation performance comparison.}
\end{figure*}

In this section, numerical simulations are conducted 
to confirm the validity and performance of the proposed approach. 
According to \cite{Wang2017CMCB}, the MRA has the maximum virtual aperture 
in the condition of the same array element among the sparse arrays; 
therefore, the delay pattern we set up
$C = \left[ {{c_1}, \ldots ,{c_M}} \right],0 = {c_1} <  \cdots  < {c_M}$ is similar 
to the array element position of MRA. 
In our simulation we choose the Nyquist sampling rate 
$f_{nyq}={{\rm{1}} \mathord{\left/{\vphantom {{\rm{1}} T}} \right.
		\kern-\nulldelimiterspace}T} = 10\text{GHz}$,  the minimal delay unit $\tau  = T$,
the number of paths following each array element $M = 4$,
and the delay pattern  $C =[ {0,1,4,6}]$.
Hence, the path number of virtual time-delay channels of delay pattern $C$
i.e., the dimension of expansion,
is $Q = 7$. 
The Root Mean Square Error (RMSE) of parameters is defined as
$\text{RMSE} = \sqrt {\frac{1}{{{N_m}K}}\sum_{i = 1}^{{N_m}} {\sum_{k = 1}^K {{{\left( {u_k^i - \hat u_k^i} \right)}^2}}}}$ 
where the superscript $i$ refers to the $i^{th}$ trail, 
and $N_m$ denotes the number of Monte Carlo tests. 
$u_k^i $ and $\hat u_k^i$ are the true parameter and estimated parameter in the $i^{th}$ trail, respectively.
The Signal to Noise Ratio (SNR) is defined as
$\text{SNR} = ( {{{E( {{{\left| {{x_m}\left( t \right)} \right|}^2}} )} \mathord{/
			{\vphantom {{E( {{{\left| {{x_m}\left( t \right)} \right|}^2}} )} {{\sigma ^2}}}}
			\kern-\nulldelimiterspace} {{\sigma ^2}}}} )$.

 In the first simulation, the aim is to verify 
 whether the algorithm JDF4BA with ETM method is valid 
 and whether the conclusion of \ref{Analyses} is correct. 
 We set up 6 QPSK signals 
of which the maximum information bandwidth inside is $B = 25\text{MHz}$
with following different carrier frequencies (in GHZ) 
$\left\{ {1.22,\;2.77,\;4.32,\;6.54,\;7.64,\;8.48} \right\}$ 
and corresponding DOAs (in degrees) 
$\left\{ {45,\;20,\;0,\; -30,\;10,\; - 20} \right\}$. 
We choose the sub-Nyquist sampling rate $f_{sub} = 25 \text{MHz}$, 
same as the maximum information bandwidth $B$,
corresponding to sampling rate reduction factor  $L=400$.
As shown in Fig.2 (a), according to the \eqref{eq-Pmusic(f)},
the persudo spectrum of frequency has 6 peaks 
so that we can obtain frequency estimation of signals.
Seeing Fig. 2 (b), we then use \eqref{eq-Pmusic(phi)} to get 6 persudo spectrums of DOA,
combined with $\{ {{{\hat f}_k}}\}_{k = 1}^K$. 
Finally, we can obtain $\{ {{{\hat f}_k},{{\hat \theta }_k}} \}_{k = 1}^K$.
According to simulation settings,
the total sampling rate $f_{sub}^{total} = 2M \cdot f_{sub}=200 \text{MHz}$ is less 
than the total Nyquist sampling rate $f_{nyq}^{total} = 2M \cdot f_{nyq} = 80 \text{GHz}$.

The second numerical simulation is 
to test the performance of algorithm JDF4BA with ETM method with different SNRs.
We set up 6 complex sinusoid signals with following carrier frequencies 
$\left\{ {1.22,\;2.77,\;4.32,\;6.54,\;7.64,\;8.48} \right\}$  
and corresponding DOAs 
$\left\{ {10,\;20,\;30,\;30,\;50,\;80} \right\}$. 
We choose the sub-Nyquist sampling rate $f_{sub} = 250 \text{MHz}$, 
corresponding to sampling rate reduction factor  $L=40$.
The number of Monte Carlo is 200.
The conditions for comparison are the algorithm JDF4BA with ETM method at the Nyquist sampling rate,
and the algorithm JDF4BA without ETM method at the sub-Nyquist sampling rate as well as the Nyquist sampling rate. 
Due to algorithm JDF4BA without ETM method, the number of identifiable signals is $M-1=3$. 
Therefore, in the numerical simulation of the algorithm without ETM method,
the signals we set up is the first three signals in set $\left\{ {{f_k},{\theta _k}} \right\}_{k = 1}^K$.

As shown Fig.2 (c) (d), we can see the performance of frequencies and DOAs via JDF4BA with ETM 
is not as good as the performance of only JDF4BA under the same sampling rate. The reason is that ETM method is in exchange for slight sacrificing performance to be able to identify more signals. Moreover, only if the sub-Nyquist sampling rate $f_{sub}$ is larger than the maximum bandwidth $B$,
the performance is almost identical to that at the Nyquist sampling rate, 
because the information bands of signals are complete at sub-Nyquist sampling. Therefore, the performance is not related with the sampling rate if the sampling rate is larger than the maximum bandwidth.
 
\section{Conclusion}
In this paper, we put forward 
a binary array radar structure (each array element has $M$ branches)
and a novel approach (algorithm JDF4BA and ETM method) 
to jointly estimate frequencies and corresponding DOAs at sub-Nyquist sampling. 
Algorithm JDF4BA can achieve auto-pairing between frequencies and corresponding DOAs
under the sub-Nyquist sampling rate $f_s \geq B$, 
thus avoiding ambiguity. 
Furthermore, ETM method increases the number of identifiable signals from $M-1$ to $Q-1$.
As a consequence, the binary array radar structure with this approach 
just needs $2M$ paths to estimate $Q-1$ signals 
with the total sampling rate $f_{sub}^{total} = 2MB$,
which means fewer number of array elements (only $2$),
fewer number of channels,
and smaller total sampling rate under the same number of signals, 
compared with the previous works. 
Simulation results validate the effectiveness and performance of the present approach.

\section*{Acknowledgment}

This work was supported by the Fundamental Research Funds
for the Central Universities (Grant No. ZYGX2016Z005, and ZYGX2016J218).

\end{document}